# A Preliminary Study on Automatic Motion Artifact Detection in Electrodermal Activity Data Using Machine Learning*

Md-Billal Hossain, *Member, IEEE*, Hugo F. Posada-Quintero, *Member IEEE*, Youngsun Kong, *Member, IEEE,* Riley McNaboe and Ki H. Chon, *Senior Member, IEEE*

*Abstract*— The electrodermal activity (EDA) signal is a sensitive and non-invasive surrogate measure of sympathetic function. Use of EDA has increased in popularity in recent years for such applications as emotion and stress recognition; assessment of pain, fatigue, and sleepiness; diagnosis of depression and epilepsy; and other uses. Recently, there have been several studies using ambulatory EDA recordings, which are often quite useful for analysis of many physiological conditions. Because ambulatory monitoring uses wearable devices, EDA signals are often affected by noise and motion artifacts. An automated noise and motion artifact detection algorithm is therefore of utmost importance for accurate analysis and evaluation of EDA signals. In this paper, we present machine learning-based algorithms for motion artifact detection in EDA signals. With ten subjects, we collected two simultaneous EDA signals from the right and left hands, while instructing the subjects to move only the right hand. Using these data, we proposed a cross-correlation-based approach for non-biased labeling of EDA data segments. A set of statistical, spectral and model-based features were calculated which were then subjected to a feature selection algorithm. Finally, we trained and validated several machine learning methods using a leave-one-subject-out approach. The classification accuracy of the developed model was 83.85% with a standard deviation of 4.91%, which was better than a recent standard method that we considered for comparison to our algorithm.

## I. INTRODUCTION

Electrodermal activity (EDA) represents changes in electrical conductance of the skin due to opening of sweat pores [1], [2]. It is believed that EDA represents sudomotor activities, as they are innervated by sympathetic C nerve fibers of the autonomic nervous system. Therefore, EDA has the potential to be used for the evaluation of sympathetic nervous function and cognitive arousal [3]–[7]. Due to ease of use and noninvasive data collection, EDA has been applied in many diverse areas including emotional arousal [8], decision-making [9], pain [10], [11], stress [12], autism [13], and panic disorder [14], as these are all related to elevation of sympathetic nervous activities.

While EDA has potential to be used as a peripheral sympathetic nervous activity marker, it has some limitations. For example, EDA can be affected by non-sympathetic factors such as atmospheric temperature and humidity [15]. Most of the traditional EDA studies were performed carefully in a controlled laboratory setting with limited movement of the subjects in order to minimize the effect of motion artifacts and noise. However, recently there have been a large number of studies of ambulatory EDA data collection over long time periods using wearable devices [7]. Wearable devices are more prone to noise and artifacts [16] and EDA is not immune from this issue. Noise and motion artifacts can be generated from different sources such as poor contact between skin and the recording electrodes, movements that cause variations in the skin-electrode contact, intentional or unintentional touching of electrodes (e.g. autistic people might be prone to the latter), and contextual factors (e.g. temperature and humidity) that may cause excessive sweating. Therefore, for accurate evaluation of EDA signals, corrupted segments should be automatically identified and removed.

Despite an increasing volume of EDA research over the last decades, there has been only a handful of research papers on motion artifact detection in EDA. Many researchers used either exponential smoothing [17] or low pass filtering [18] to combat noise and motion artifacts. These techniques may smooth high variations in the signal; however, they cannot compensate for sudden and large-magnitude motion artifacts which are often present in EDA signals, especially in ambulatory recordings. There have been some heuristic methods [18] to remove motion artifacts in EDA. However, all the methods were developed on a specific dataset, hence, their performance becomes ineffective for untrained datasets. Taylor et al. [19] developed a machine learning algorithm using the support vector machine (SVM) classifier. They have used manually annotated EDA segments and extracted different statistical features to train an SVM classifier. Ian et al. [20] proposed a simple EDA quality assessment procedure based on some simple decision rules. While this method works for spiky and large amplitude motion artifacts, it fails in several other cases.

It should be noted that unlike other biosignals, such as electrocardiogram (ECG) and photoplethysmography (PPG), EDA does not exhibit periodicity. Hence, manual adjudication of clean versus noisy EDA can be rather tricky. In order to avoid this issue, we propose an automated correlation-based annotation criterion to determine if an EDA data segment is clean or noisy. Certainly, this is not a practical solution since this scheme will require a stationary reference signal. However, for our purpose in this work, this approach was used to determine clean versus noisy data. Finally, based on the adjudicated data from the cross-correlation scheme, we developed a machine learning algorithm to identify the noisy EDA segments using several statistical and model-based features. This machine learning is the ultimate outcome of an

*Research supported by Office of Naval Research.
MB Hossain, H.F. Posada-Quintero, Y. Kong, R. McNaboe and Ki H. Chon are with the Department of Biomedical Engineering at University of Connecticut, Storrs, CT 06269, USA. email: {md.b.hossain, hugo.posada-quintero, youngsun.kong,riley.mcnaboe,ki.chon} at uconn.edu

automated way of adjudicating clean versus noisy data in practice.

## II. METHODS AND MATERIALS

### A. Data Collection

Ten healthy subjects, aged 20-35, participated in this study. Two channels of simultaneous EDA were collected from the right and left hands using ADInstrument's galvanic skin response modules. On each hand, a pair of stainless-steel electrodes were placed on index and middle fingers. The EDA signals were recorded at a sampling frequency of 1,000 Hz, then down-sampled to 8 Hz. The data collection protocols were designed so that the right hand made occasional movements to mimic regular motion artifacts people could create in their daily lives while the left hand was immobile in order to provide a reference EDA. All the experimental procedures were approved by the Institutional Review Board (IRB) for human subject research at The University of Connecticut. A summary of the protocol is shown in Table I. The protocol consists of two parts, where the first part was done with no significant motion and the second part was designed to mimic different motion artifacts.

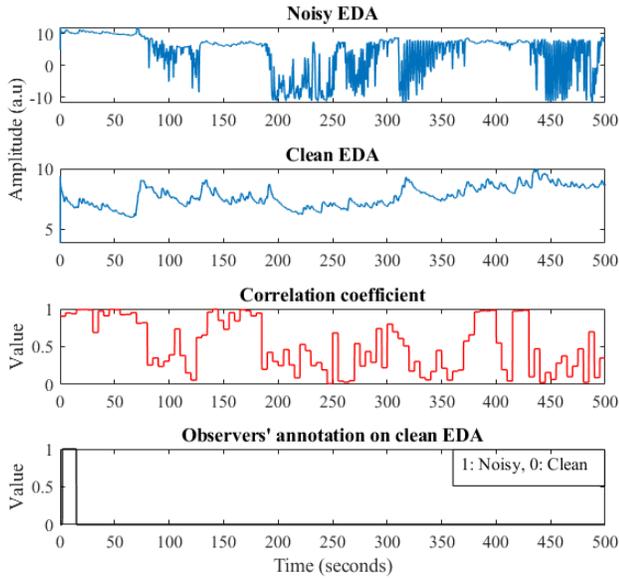

Figure 1. Noisy EDA channel (right hand), clean or reference EDA (left hand), correlation coefficient and observers' annotation of reference EDA.

### B. Data Labeling

As mentioned earlier, EDA data quality labeling can be difficult given the non-conventional characteristics of EDA signals, hence, observers' adjudications can be incorrect. To avoid human adjudication bias, we propose a cross-correlation coefficient-based criterion to annotate EDA segments as either noisy or clean. EDA data were first segmented into 1692 non-overlapping 5-second windows. We computed the cross-correlation coefficient between each of the simultaneous reference (from the left hand) EDA segments and the targeted EDA segments (from the right hand). If the correlation coefficient exceeded 0.85, we considered this targeted EDA segment as clean, otherwise it was considered noisy. Please note that the reference signal can be noisy occasionally because of intentional or unintentional movements. Therefore, we first manually checked the reference and marked the obviously noisy portions of EDA. For this purpose, we used three independent reviewers' annotations; any segment that was marked as noisy by any of the three reviewers was discarded from the analysis unless there was a very high correlation ($> 0.95$) between the reference and the target EDA segment. Fig. 1 shows an example of two simultaneous EDA channels (target EDA and reference EDA), their corresponding cross-correlation coefficient, and the observers' annotation for the reference EDA. It can be seen from Fig.1 that the correlation coefficient follows the target EDA, meaning that when there is noise in the target EDA the correlation coefficient is low, and vice versa.

TABLE I. DATA COLLECTION PROTOCOL SUMMARY

| Duration (second) | Activity | Remarks |
|---|---|---|
| Part I (Stress Test) | | |
| 120 | Flat table, relaxing with eyes closed | Baseline |
| 30 | Start table tilt | Orthostatic Stress |
| 120 | Subject remains in tilted position | |
| 150 | Return table to flat position, subject relaxation | |
| 120 | Perform Stroop test | Cognitive Stress |
| Part II (Motion Artifact Test) | | |
| 60 | Sitting up in a chair with no movement | Motion Artifact Induction |
| 60 | Sitting down and typing on the computer | |
| 60 | Sitting down and holding a mouse, clicking the mouse | |
| 60 | Standing up with the arm next to torso | |
| 60 | Standing up swinging arm by the side as if to simulate walking | |
| 60 | Standing up with arm straight out, moving arm up and down continuously | |
| 60 | Standing up with arm straight out moving the arm at the elbow allowing the wrist to come into the chest and straight out and do the same again. | |
| 60 | Standing up with arm straight, completing a circle with the fingertips by rolling the shoulder. | |

### C. Feature Extraction

We computed a set of features from each of the 5 second EDA window segments. First, we computed different statistical features from the original EDA, and its first and second derivative. We modeled the EDA sequence using an autoregressive (AR) model and considered 2 AR parameters ($a_1$ and $a_2$) and the AR noise variance as features. The motivation behind including AR modelling is that when EDA data are corrupted by noise, there will be more residual noise in the AR model than in the clean data; similarly, both AR parameters will have greater values for the noisy than clean data.

We then used a high-resolution time frequency decomposition method called variable frequency complex demodulation (VFCDM) [21], which provides more dynamic features of the clean and corrupted EDA. VFCDM has been previously used for several biosignal applications and has been found to be effective in analyzing signal characteristics [22] and removing noise and artifacts [23], [24]. We decomposed the EDA data segments into 12 non-overlapping frequency bands using VFCDM. Finally, we reconstructed two signals using the first 3 modes of VFCDM for the first signal, and the rest of the modes for the second signal. We computed the mean, variance, ratio of the variances, and ranges (max-min) of these two signals using VFCDM. Note that modes (1-3) include most of the dynamic characteristics of the EDA. Therefore, these features may reflect the signal and noise strength in the EDA segment.

As in [19], we also used the discrete wavelet transform and computed several other features from the details and approximation coefficients. We used three-level wavelet decomposition using the Haar wavelet. By doing so, we may have included some redundant features. However, we were not concerned about including too many features because we used a feature selection algorithm to reduce the risk of overfitting, as described in the subsequent section. All features computed are shown in Table II.

TABLE II. SUMMARY OF THE FEATURES COMPUTED

| Index | Category | Specific features |
|---|---|---|
| 1-3 | AR(2) Modelling | AR parameters, AR noise variance |
| 4-9 | Raw EDA | Mean, median, variance, Shannon entropy, range, and skewness |
| 10-19 | First and second derivative of EDA | Mean, variance, max, and min of the absolute value |
| 20-43 | Wavelet decomposition | Mean, median, variance, Shannon entropy, range, and number above zero of the wavelet coefficients |
| 44-52 | VFCDM decomposition | Mean, variance, range of the two intermediate reconstructed signals, and ratio of the variances. |

### D. Feature Selection and classification:

We used the random forests (RF) machine learning algorithm for feature selection [25]. RF is a popular feature selection algorithm because of its good predictive performance, low overfitting, and interpretability. Feature selection using RF is in the embedded methods category which is a fusion of filter and wrapper methods. The embedded methods are highly accurate, easily generalizable, and interpretable in terms of feature selection.

To perform feature and model selection we followed a subject-independent validation strategy. We performed a leave one subject out (LOSO) validation strategy, meaning in every fold we left one subject out for testing and used the rest for training. We tested several classifiers such as random forests (RF), support vector machine (SVM) with linear and radial basis function (RBF) kernels, and K nearest neighbor (KNN). Among those tested, RF and SVM with RBF kernel showed the best performance. We obtained almost similar results using both RF and SVM with RBF kernel, hence, we present the results for only these two classifiers. For feature selection we used a group k-fold validation using the training data at each fold of LOSO validation. Again, we used a group k-fold to ensure the classifiers were subject-independent. To select the optimal parameter for each fold, we performed a grid-search cross-validation technique with the group k-fold. The parameters $C$ and gamma for SVM were selected from parameter candidates of 1, 10, 100, and 1000, and 0.001, 0.01, 0.1, 1, respectively.

### III. RESULTS

#### A. Classification results

Table III shows the classification results using both RF and SVM. We compared the classification performance with a previously published motion artifact detection algorithm, which showed promising results [20].

TABLE III. CLASSIFICATION RESULTS AND COMPARISON

| Methods | Mean accuracy | Standard deviation |
|---|---|---|
| Ian et al. [20] | 75.05% | 9.73% |
| This work (RF) | 83.40% | 4.06% |
| This work (SVM) | 83.85% | 4.91% |

As shown in Table III, SVM provided the highest detection accuracy. The performance of [20] is lower compared to SVM. The performance of RF and SVM was similar; SVM had slightly higher accuracy (83.85%) albeit at the expense of higher standard deviation. While RF and SVM have standard deviation less than 5% (4.06% and 4.91%, respectively), [20] showed a standard deviation of 9.73%. This suggests that the machine learning methods used in this work are more consistent for motion artifact detection.

#### B. Feature Analysis

We have tracked the features selected by RF at each fold and the most consistent features that were selected are presented in Table IV. As mentioned earlier, at every fold we performed a group five-fold validation on the training data to select the best feature combinations. For the RF classifier, the features selected did not change the classification accuracy significantly, however, for SVM the accuracy increased by around 3.2% (SVM accuracy was 80.65% before feature selection). Finally, we have statistically analyzed some of the features selected by the RF feature selection. We performed a pairwise t-test for the feature values of the two classes and found statistically significant difference in most of the selected features.

TABLE IV. FEATURES SELECTED CONSISTENTLY

| Category | Features selected |
|---|---|
| AR modeling | Noise variance, AR parameters ($a_1$ and $a_2$) |
| Raw EDA | Mean, range |
| | 1st derivative: max |
| | 2nd derivative: range |
| VFCDM | Variance in first three modes and rest of the modes (4-12) |
| Wavelet | Mean of approximations |
| | Variances of details coefficients |

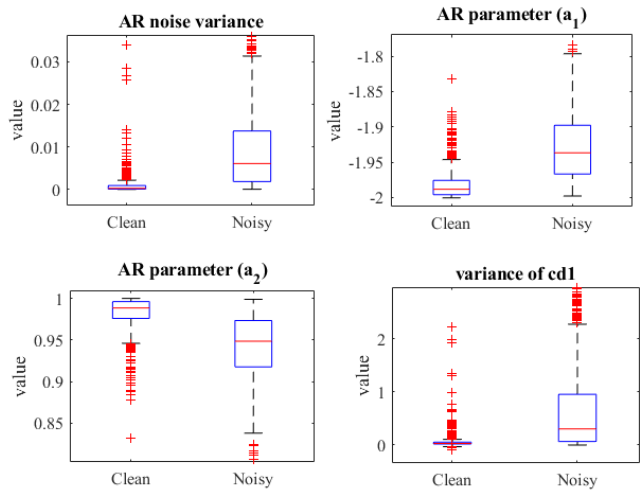

Figure 2. Selected feature statistics for clean and noisy EDA.

Fig. 2 shows box plots of the distribution of four of the consistent features' distributions for the clean and noisy EDA. As shown in Fig. 2, there is a clear visual difference between the mean values of the features of two classes. As expected, the AR model had higher residual noise for the noisy EDA; the AR parameters also reflect this observation. The same is noted for the variance of cd1 (detail coefficients 1). The t-test also provided statistically significant differences ($p < 0.01$).

## IV. CONCLUSIONS

We presented preliminary results of an automated motion artifact detection algorithm for EDA signals. We created an EDA database with simultaneous reference and target EDA channels and proposed a correlation-based criterion to define clean and noisy EDA segments without human annotations which could be biased and may differ from person to person. We have proposed several statistical, model-based and time-frequency features and applied a subject-independent machine learning algorithm for automated motion artifact detection in EDA signals. We also selected the consistent features using RF for subsgequent classification analysis. The performance of the machine learning-based motion artifact detector is compared with a recently published promising method. It was observed that our approach performed better and with more consistency. While the proposed method showed promising performance, more data and further rigorous analysis is needed in the future to confirm the results.